\documentclass[a4,semcolor,psfig,english,12pt,epsf,portrait]{article}
\usepackage{amsmath,amsfonts,latexsym,amscd,amssymb,theorem}

\usepackage[T1]{fontenc}
\usepackage{epsfig}
\usepackage{amsmath}
\usepackage{psfrag}

\def\R{\hbox{\bf R}}
\def\Z{\hbox{\bf Z}}

\def\T{\mathbb{ T}}

\def\<{\langle}
\def\>{\rangle}
\newcommand{\ba}{\begin{eqnarray}}
\newcommand{\ea}{\end{eqnarray}}

\newtheorem{theo}{\bf Theorem}[section]

\newtheorem{rem}[theo]{\bf Remark}

\renewcommand{\R}{{\mathbb R}}
\renewcommand{\Z}{{\mathbb Z}}

\begin{document}

\title{\bf Derivation and study of dynamical models\\ of dislocation
  densities }

\author{
\normalsize\textsc{ A. El
  Hajj$^1$}, H. Ibrahim$^2$, R. Monneau$^2$}
\vspace{20pt}
\maketitle
\footnotetext[1]{Universit\'e d'Orl\'eans,
Laboratoire MAPMO,
Route de Chartres, 45000 Orléans cedex 2, France}
\footnotetext[2]{Ecole Nationale des Ponts et Chauss\'ees, CERMICS,
 6 et 8 avenue
  Blaise Pascal Cit\'e Descartes Champs-sur-Marne, 77455
  Marne-la-Vall\'ee  Cedex 2, France}


 \centerline{\small{\bf{Abstract}}}
 \noindent{\small{In this paper, starting from the microscopic dynamics
    of isolated dislocations,  
we explain how to derive formally mean field models for the dynamics
of dislocation densities. Essentially these models are tranport equations,
coupled with the equations of elasticity. Rigorous results of existence of solutions are
presented for some of these models and the main ideas of the proofs are
given.}}

\hfill\break
 \noindent{\small{\bf{AMS Classification: }}} {\small{54C70, 35L45, 35Q72, 74H20, 74H25.}}\hfill\break
  \noindent{\small{\bf{Key words: }}} {\small{Cauchy's problem,
      non-linear transport equations,  non-local transport 
equations,  hyperbolic equations, BMO estimate, 
dynamics of dislocation densities.}}\hfill\break

\section*{Introduction}
In this paper we are interested in several mesoscopic models involving the
dynamics of dislocations. These models are important for the understanding
of the elasto-visco-plasticity  behaviour of materials. For crystals, at the
microscopic level, the origin of plasticity is mainly due to the existence
of defects called dislocations. Dislocations are line defects that can move in
the crystal when a shear stress is applied. On the other hand each
dislocation also creates a stress field. This leads to a complicated
dynamics of these defects that we consider below. See Figure \ref{f0} for
an example of complicated pattern of dislocations in real crystals.\\

\begin{figure}[h]\label{f0}
\centering\epsfig{figure=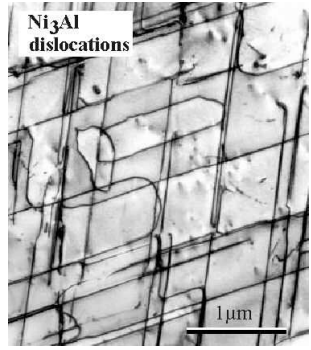,width=40mm}
\caption{Dislocation lines in Nickel Aluminium alloy.}\label{f0}
\end{figure}

\noindent {\bf  Organization of the paper.}\\
Section \ref{s1} presents the microscopic modelling for the dynamics of
dislocation straight lines. In Section \ref{s2}, we explain how to derive
formally a two-dimensional mean field model, called the Groma, Balogh
model. A rigorous result of existence of solutions is given and other
results are also given for a one-dimenional submodel. In Section \ref{s3},
the Groma, Czikor, Zaiser model is studied. In particular is presented a simulation of
the deformation of a slab under an external shear stress. In Section
\ref{s4}, the formal derivation of a mean field model for densities of
dislocation curves is presented. Concluding remarks are given in Section
\ref{s5}.

\section{The microscopic modelling}\label{s1}

\subsection{Preliminary: the stress field created by an edge dislocation}

In the space coordinates $x=(x_1,x_2,x_3)$ with basis $(e_1,e_2,e_3)$, 
we consider the case a material filling the whole space and with a
single dislocation line which is the 
axis $x_3$. To this line is associated an invariant, called the Burgers
vector $b$. In the case of linear isotropic elasticity with Lam\'e
coefficients $\lambda,\mu$ (satisfying $\mu>0, 3\lambda + 2\mu >0$),
the stress created by the dislocation is given by 
\begin{equation}\label{eq::sigmadefi}
\sigma_{ij}=2\mu e_{ij} + \lambda \left(\sum_{k=1,2,3} e_{kk}\right)
\delta_{ij},\quad i,j=1,2,3   
\end{equation}
where
$$e_{ij} = \frac12\left(\frac{\partial u_j}{\partial x_i}+\frac{\partial
  u_i}{\partial x_j}\right) + H(x_1)\delta_0(x_2) e^0_{ij} \quad 
\mbox{with} \quad e^0=\frac12\left(b\otimes n + n\otimes b\right)$$
where $H$ is the Heaviside function, $\delta_0$ is the Dirac mass,
$n=e_2$ is the normal to the slip plane  
and $(u_1,u_2,u_3)$ is the three-dimensional displacement field.
In the special case where $b=e_1$, the dislocation is called an edge
  dislocation (the Burgers vector is perpendicular to the dislocation
  line). In this case we have
\begin{equation}\label{eq::e0}
e^0=\frac12 \left(\begin{array}{lll}
0&1&0   \\
1&0&0\\
0&0&0
\end{array}\right).   
\end{equation}
The stress is assumed to satisfy the elasticity equation of equilibrium
\begin{equation}\label{eq::sigma}
\sum_{i=1,2,3}\frac{\partial \sigma_{ij}}{\partial x_i}=0,\quad j=1,2,3. 
\end{equation}
The solution to this equation is known since the work of Volterra
\cite{V}. We have (see Hirth, Lothe \cite{HL}):
\begin{equation}\label{eq::s}
\sigma(x) = a \left(\begin{array}{ccc}
\displaystyle{-\frac{x_2(3x_1^2+x_2^2)}{(x_1^2+x_2^2)^2}}  &
\displaystyle{\frac{x_1(x_1^2-x_2^2)}{(x_1^2+x_2^2)^2}} & 0\\
\\
\displaystyle{\frac{x_1(x_1^2-x_2^2)}{(x_1^2+x_2^2)^2}} &
\displaystyle{\frac{x_2(x_1^2-x_2^2)}{(x_1^2+x_2^2)^2}} & 0\\
\\
0 & 0 & 0
\end{array}\right)   
\end{equation}
with 
$$a=\frac{\mu}{2\pi (1-\nu)}>0 \quad \mbox{where the Poisson ratio is}
\quad \displaystyle{\nu=\frac{\lambda}{2(\lambda +\mu)}\in \left(-1,\frac12\right)}.$$

\subsection{A two-dimensional microscopic model}

In this section we describe very formally a microscopic model describing
the dynamics of dislocation lines in the particular geometry where all
lines are parallel to the axis $x_3$.\\

\begin{figure}[h]
\centering\epsfig{figure=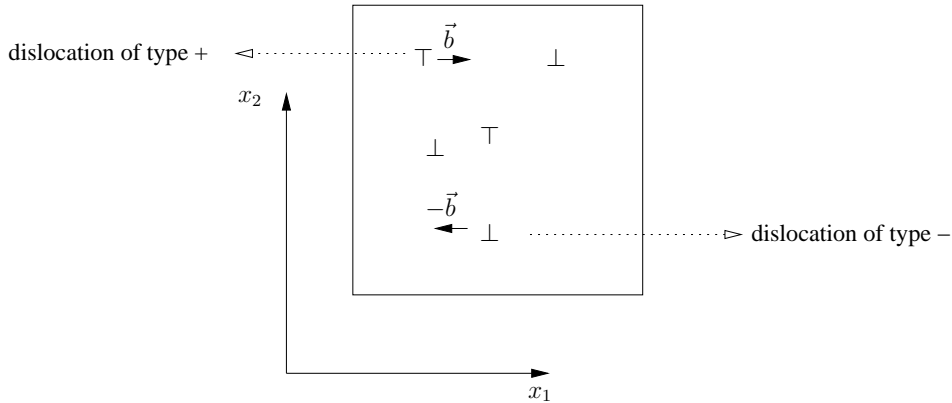,width=130mm}
\caption{The cross section of the dislocation lines.}\label{f2} 
\end{figure}

Considering the cross section of these
lines, we can reduce the problem to a two-dimensional problem where each
dislocation line can be identified to its position $(x_1,x_2)$. For
$i=1,...,N$, let us call $X^{i}$ the positions in $\R^2$
of these dislocations (see Figure \ref{f2}).

We consider moreover the particular case where each dislocation can only move along the
direction $e_1$. Those dislocations $X^i$ are known to be characterized by an
invariant called the Burgers vector $b_i=\varepsilon_i e_1$ where
$$\varepsilon_i=\pm 1.$$
Finally the dynamics of these dislocations is
\begin{equation}\label{eq::0}
\frac{dX^i(t)}{dt} =b_i\ \sigma_{12}(X^i(t)),\quad \mbox{for}\quad i=1,...,N.  
\end{equation}
Here $\sigma_{12}$ is one component of the stress. In the case of linear
isotropic elasticity, it is known that
$$\sigma_{12}(x)= \sum_{i=1,...,N} \varepsilon_i \sigma_0(x-X^i),\quad \mbox{where}\quad
\sigma_0(x)=a \ \displaystyle{\frac{x_1(x_1^2-x_2^2)}{(x_1^2+x_2^2)^2}},$$
where $\sigma_0$ is the $(1-2)$-component of the stress given in (\ref{eq::s}).
Here $\sigma_0(x)$ appears to be the shear stress created at the point $x$
by a single dislocation positionned at the origin, with Burgers vector
$e_1$.
With the convention that the self-stress created by a dislocation on itself
is zero, i.e. formally $\sigma_0(0)=0$, we see that we can rewrite the full dynamics
of particles satisfying $X^i\not= X^j$ for $i\not= j$,
as follows
$$\frac{dX^i}{dt} =e_1\ \left(\sum_{j\in \left\{1,...,N\right\}\backslash
  \left\{i\right\}} \varepsilon_i\varepsilon_j
  \sigma_0(X^i-X^j)\right),\quad \mbox{for}\quad i=1,...,N.$$
Let us now introduce the ``density'' of dislocations
$$\theta^\pm(x,t)= \sum_{i\in \left\{1,...,N\right\}\ :\ \varepsilon_i=\pm 1}\delta_0(x-X^{i}(t))$$
where $\theta^+$ and $\theta^-$ correspond respectively to the dislocations with
positive and negative Burgers vector. Again with the convention
$\sigma_0(0)=0$, we can rewrite the dynamics as
\begin{equation}\label{eq::1}
\theta^\pm_t + \mbox{div}\ \left(\pm J\ \theta^\pm \right) =0\quad \mbox{with}\quad J=e_1\ \sigma_0\star
  (\theta^+-\theta^-).   
\end{equation}

\section{Mesoscopic models}\label{s2}

\subsection{Formal derivation of the two-dimensional mesoscopic model}

A natural question is what is the effective
model at the mesocsopic scale corresponding to the microscopic model
(\ref{eq::1})?
It seems to be an open and difficult question. As a naive answer, we can
consider as a first candidate to the mesoscopic model, the mean field model
corresponding to the model (\ref{eq::1}), where now each density $\theta^+$
and $\theta^-$ can be seen as a ``continuous'' quantity. In this mean field
model, we have in particular neglected the short range dynamics. One convenient way
to rewrite the model is to introduce one primitive $\rho^\pm$ of the densities
$\theta^\pm$ such that
$$\rho^\pm_{x_1}(x,t)=\theta^{\pm}(x,t) \ge 0.$$
Then we can rewrite (\ref{eq::1}) as (with a zero constant of integration):
\begin{equation}\label{eq::2}
\rho^\pm_t = \mp \sigma_{12}\ \rho^\pm_{x_1}
\end{equation}
with
\begin{equation}\label{eq::3}
\sigma_{12}= (\sigma_0)_{x_1} \star (\rho^+-\rho^-). 
\end{equation}
System (\ref{eq::2})-(\ref{eq::3}) is an integrated form of the Groma,
Balogh model (\ref{eq::1}) (see \cite{GB}).
It turns out that by construction $\sigma$ solves: 
\begin{equation}\label{eq::9}
\left\{\begin{array}{l}
\displaystyle{\sum_{i=1,2,3}\frac{\partial \sigma_{ij}}{\partial x_i}=0},   \\
\displaystyle{\sigma_{ij}=2\mu e_{ij} + \lambda \left(\sum_{k=1,2,3} e_{kk}\right)
\delta_{ij}},\\
\displaystyle{e_{ij} = \frac12\left(\frac{\partial u_j}{\partial x_i}+\frac{\partial
  u_i}{\partial x_j}\right) + (\rho^+-\rho^-) e^0_{ij}}, 
\end{array}\right.   
\end{equation}
where  $e^0$ is given in (\ref{eq::e0}). Here (\ref{eq::3}) appears to be a
representation formula for the solution of the equations of elasticity
(\ref{eq::9}).

\subsection{Existence of a solution in the two-dimensional periodic case}

In the case where the dislocations densities $\theta^\pm(x,t)$ are
$\Z^2$-periodic in $x \in\R^2$ such that 
$$\int_{0}^1 dx_1 \ \theta^\pm(x_1,x_2,t)=L>0,$$
 it turns out that the representation formula
(\ref{eq::3}) can be rewritten
\begin{equation}\label{eq::4}
\sigma_{12}=\bar{a}\  R_1^2R_2^2 (\rho^+-\rho^-)\quad \mbox{with} \quad
\bar{a}=4\mu(\lambda+\mu)/(\lambda +2\mu),  
\end{equation}
where $R_i$ for $i=1,2$ are the Riesz transform given in Fourier series
coefficients on $\T^2=\R^2 / \Z^2$  by
$$c_k(R_i(f))=\left\{\begin{array}{ll}
\displaystyle{\frac{k_i}{|k|} c_k(f)} & \quad
\mbox{for all}\quad k=(k_1,k_2)\in\Z^2\backslash \left\{(0,0)\right\}\\
0 & \quad \mbox{if}\quad k=(0,0)
\end{array}\right.$$
where
$$c_k(f)=\int_{\T^2} e^{-2i\pi k\cdot x} f(x) \ d^2x.$$
We will also make use of the norm on the following Zygmund space
$$||f||_{L\ln L(\T^2)}= \inf\left\{\gamma >0,\quad \int_{\T^2}
  \frac{|f|}{\gamma} \ln\left(e +\frac{|f|}{\gamma}\right)\le 1\right\}.$$
We assume that
\begin{equation}\label{eq::8}
\rho^\pm(\cdot,0)=\rho^\pm_0   
\end{equation}
and  we make the following assumptions on the initial data
\begin{equation}\label{eq::5}
\left\{\begin{array}{l}
\rho^\pm_0(x_1 + m,x_2 +l)=mL + \rho^\pm_0(x_1,x_2) \quad \mbox{for
  any}\quad m,l\in\Z\\
\\
(\rho^\pm_0)_{x_1}\ge 0\\
\\
||(\rho^\pm_0)_{x_1}||_{L\ln L(\T^2)} \le C 
\end{array}\right.
\end{equation}
where $C>0$ is any fixed constant.\\

Then we have:
\begin{theo}\label{th:1}{\bf (Global existence of a solution in the
    periodic case, \cite{CEMR})}\\
Under assumption (\ref{eq::5}) there exists a function $(\rho^+,\rho^-)$
which is a solution of
(\ref{eq::2}), (\ref{eq::4}) in the sense of distributions and with initial
condition (\ref{eq::5}), such that $\rho^\pm(\cdot,t)$ satisfies
(\ref{eq::5}) for all time. Moreover $\rho^\pm \in
C([0,+\infty);L^1_{loc}(\R^2))$ and 
\begin{equation}\label{eq::6}
\sigma_{12}\in L^2_{loc}([0,+\infty);H^1_{loc}(\R^2)).   
\end{equation}
\end{theo}

\begin{rem}\label{}
Recall that there exists a constant $C>0$ such that the general
inequality holds for functions $f\in H^1(\T^2)$ and $g\in L\ln L(\T^2)$
$$||fg||_{L^1(\T^2)}\le C ||f||_{H^1(\T^2)} ||g||_{L\ln L(\T^2)}.$$   
Then the product $\sigma_{12}\cdot  (\rho^\pm)_{x_1}$ in (\ref{eq::2}) is
defined because of both estimate (\ref{eq::6}) on $\sigma_{12}$ and
estimate $L^\infty([0,+\infty );L\ln L(\T^2))$ on $(\rho^\pm)_{x_1}$.
\end{rem}

Indeed, in the proof of Theorem \ref{th:1}, the main tool is to consider the  entropy 
$$S(t)= \sum_{\pm} \int_{\T^2} \theta^\pm(\cdot,t) \ln \theta^\pm(\cdot,t)
\quad  \mbox{with}\quad  \theta^\pm = (\rho^\pm)_{x_1}$$
which satisfies formally the following a priori estimate
$$S(t) +\bar{a} \int_0^t \int_{\T^2} (R_1R_2(\theta^+-\theta^-))^2 \le S(0).$$
The uniqueness of the solution remains an open problem.

\subsection{A one-dimensional mesoscopic submodel}

In this section, we present a submodel where the uniqueness of the solution
is known.\\

\begin{figure}[h]
\psfrag{a}{$\bot$}
\psfrag{b}{$\top$}
\psfrag{x}{$x_1$}
\psfrag{y}{$x_2$}
\centering\epsfig{figure=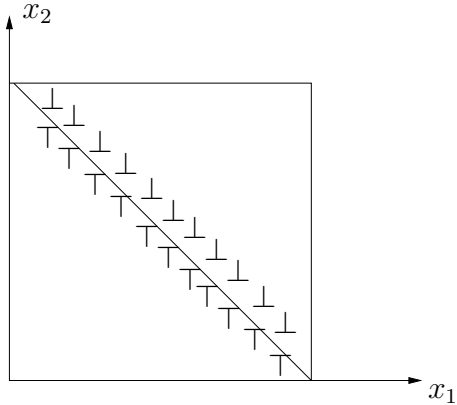,width=70mm}
\caption{A $1$-D sub-model invariant by translation in the $(-1,1)$
  direction.}\label{f3} 
\end{figure}

Let us now consider a solution $(\rho^+,\rho^-)$ of (\ref{eq::2}),
(\ref{eq::4})  only depending on the
variable $y=x_1+x_2$ and on the time $t$ (see Figure \ref{f3}). 

In that case, we can rewrite the
system of equations for $(\rho^+(y,t),\rho^-(y,t))$ as follows:
\begin{equation}\label{eq::7}
\left\{\begin{array}{l}
\displaystyle{\rho^+_t=-c_1\left\{(\rho^+-\rho^-) + c_2\int_0^1dz\
  (\rho^+(z,t)-\rho^-(z,t))\right\}\ \rho^+_y}\\
\\
\displaystyle{\rho^-_t=c_1\left\{(\rho^+-\rho^-) + c_2\int_0^1dz\
  (\rho^+(z,t)-\rho^-(z,t))\right\}\ \rho^-_y}\\
\end{array}\right. 
\end{equation}
with the constants
$$c_1=\frac{\mu(\lambda+\mu)}{\lambda +2\mu}>0,\quad
c_2=\frac{\mu}{\lambda+\mu}>0.$$
For $\rho^\pm_y\ge 0$, this system looks very much like the Burgers equation in the rarefaction
case (where no shocks are created). For this system, the following uniqueness result
is available:

\begin{theo}\label{th:2}{\bf (Uniqueness for the
    one-dimensional submodel, \cite{EF})}\\
Assume that the initial data $\rho^\pm_0$ is Lipschitz non-decreasing and
that for some $L>0$, the functions $y\mapsto \rho^\pm_0(y)-Ly$
are $1-periodic$. Then there exists a unique viscosity solution $(\rho^+,\rho^-)$
to the system (\ref{eq::7}),(\ref{eq::8}). Moreover this solution is
globally Lipschitz in space and time.
\end{theo}

In this theorem the notion of viscosity solution is the one introduced by
Ishii, Koike \cite{IK} for systems which are quasi-monotone. Roughly
speaking, this is related to the fact that for $c_2=0$, this system has a
comparison principle. More precisely, if $\rho^{\pm,1}$ and $\rho^{\pm,2}$
are two solutions of system (\ref{eq::7}) with $c_2=0$ such that 
$$\rho^{+,1}(y,t)\le \rho^{+,2}(y,t)\quad \mbox{and}\quad
\rho^{-,1}(y,t)\le \rho^{-,2}(y,t) \quad \mbox{for all}\quad y\in\R$$
holds at time $t=0$, then this is true for all time $t>0$.

Moreover for system (\ref{eq::7}), it is possible to write an upwind 
scheme and to prove a Crandall-Lions type discrete-continuous error estimate for this scheme,
as it is done in El Hajj, Forcadel \cite{EF}.

Let us mention that an existence and uniqueness result for system
(\ref{eq::7}) has also been obtained by El Hajj \cite{E} in the framework of
$H^1_{loc}(\R)$ initial data with solutions in $H^1_{loc}(\R\times [0,+\infty))$.

This system has also been studied in the case of
periodic external applied stress. In the system (\ref{eq::7}), this corresponds to add a
time-periodic term to the quantity $(\rho^+-\rho^-)$. Then for this
non-local system, it is shown formally in Briani, Cardaliaguet, Monneau \cite{BCM} (see also Souganidis,
Monneau \cite{SM} for a local system in the stationary ergodic setting)
that the long time behaviour 
of the system is an equivalent quasilinear
diffusion equation.

\section{The GCZ one-dimensional mesoscopic model}\label{s3}

\subsection{Presentation of the model}

\begin{figure}[h]
\psfrag{tau}{$\tau$}
\psfrag{1}{$1$}
\psfrag{-1}{$-1$}
\psfrag{e1}{$e_1$}
\psfrag{e2}{$e_2$}
\centering\epsfig{figure=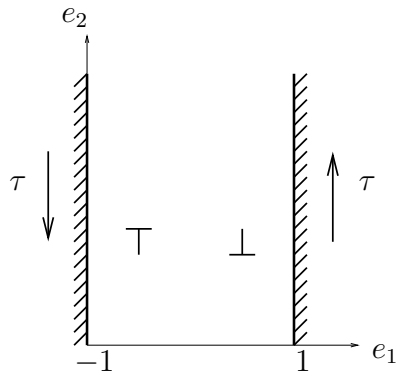,width=50mm}
\caption{Geometry of the slab.}\label{f4}
\end{figure}

In this section we consider the case where the three-dimensional material is a slab
$\Omega= (-1,1)\times \R^2$ with boundary conditions
\begin{equation}\label{eq::10}
\sigma\cdot n = \pm \tau e_2 \quad \mbox{for}\quad x_1=\pm 1   
\end{equation}
where $\tau\in\R$ is a fixed constant (see Figure \ref{f4}). 

In that case, we can check that the
solution to the equation (\ref{eq::9}) on $\Omega$ supplemented with boundary
conditions (\ref{eq::10}), is
$$\sigma_{12}=\tau \quad  \mbox{and}\quad \sigma_{ij}=0 \quad
\mbox{if}\quad \left\{i,j\right\}\not= \left\{1,2\right\}.$$
And then from the evolution equation (\ref{eq::2}), we see that in the case
$\tau>0$, the positive dislocations move to the right and the negative
dislocations move to the left. In order to take into account the short
range dynamics with accumulations of dislocations on the boundary of the
material, Groma, Czikor and Zaiser \cite{GCZ} have proposed to modify equation
(\ref{eq::2}) into the following equation (GCZ model)
\begin{equation}\label{eq::11}
\rho^\pm_t=\mp \tilde{\sigma}_{12} \rho^\pm_{x_1}   \quad \mbox{with} \quad
\tilde{\sigma}_{12} ={\sigma}_{12} + \tau_b
\end{equation}
where $\tau_b$ is the back stress created by the concentration of dislocations
$$\tau_b = -D_0\ \frac{(\theta^+-\theta^-)_{x_1}}{\theta^++\theta^-} \quad
\mbox{with}\quad \theta^\pm = \rho^\pm_{x_1}$$
where $D_0$ is a diffusion coefficient that we take equal to $1$ to
simplify the presentation.
Introducing the quantities
$$\rho=\rho^+-\rho^- \quad \mbox{and}\quad \kappa=\rho^++\rho^-,$$
the full GCZ system can be rewritten with $y=x_1\in I=(-1,1)$ as
\begin{equation}\label{eq::12}
\left\{\begin{array}{l}
\rho_t= \rho_{yy} -\tau \kappa_y\\
\kappa_t\kappa_y =\rho_t\rho_y
\end{array}\right| \quad \mbox{on}\quad I\times (0,+\infty)
\end{equation}
with boundary conditions on $\partial I$:
\begin{equation}\label{eq::13}
\left\{\begin{array}{l}
\rho(-1,t)= 0  \quad \mbox{and}\quad \rho(1,t)=0 \\
\kappa(-1,t)=-c_0 \quad \mbox{and}\quad \kappa(1,t)=c_0>0
\end{array}\right| \quad \mbox{for all}\quad t\ge 0,
\end{equation}
for some constant $c_0$
and the initial conditions
\begin{equation}\label{eq::14}
\left\{\begin{array}{l}
\rho(\cdot,0)=\rho^0   \\
\kappa(\cdot,0)=\kappa^0.
\end{array}\right.
\end{equation}
The non-negativity of the densities $\theta^\pm$ at the inital time is
equivalent to the following condition
\begin{equation}\label{eq::15}
\kappa^0_y \ge |\rho^0_y| \quad \mbox{on}\quad I.
\end{equation}
Moreover, we will assume the following condition
\begin{equation}\label{eq::16}
(\rho^0_y,\kappa^0_y) \quad \mbox{is in} \quad  C^\infty(I) \quad
\mbox{and with compact support in}\quad I.
\end{equation}

\subsection{Main results}

For this system, we have the following result 
\begin{theo}\label{th:3}{\bf (Global existence of a solution; \cite{IJM2})}\\
Assume that the initial data $(\rho^0,\kappa^0)$ satisfies
(\ref{eq::15})-(\ref{eq::16}). 
Then there exists two functions $\rho\in C^1\left(\overline{I}\times
  [0,+\infty)\right)$ and $\kappa \in C^0\left(\overline{I}\times
  [0,+\infty)\right)$ solution of
(\ref{eq::12})-(\ref{eq::13})-(\ref{eq::14}). Moreover for all time $t>0$,
(\ref{eq::15}) is satisfied with $(\rho^{0},\kappa^{0})$ replaced by
$(\rho(\cdot,t),\kappa(\cdot,t))$. 
\end{theo}

Remark that in this theorem the notion of solution to (\ref{eq::12}) is the
following. The first equation of (\ref{eq::12}) is satisfied in the sense
of distributions, while the second equation of (\ref{eq::12}) satisfied by
$\kappa$ is satisfied in the viscosity sense. This makes sense because
the right hand side $\rho_t\rho_y$ of the equation is continuous, and
because the time derivation $\kappa_t$ is multiplied by the  non-negative
quantity $\kappa_y$.

Here condition (\ref{eq::16}) appears to be a technical condition to get
the result. Althought this one-dimensional system (\ref{eq::12}) seems very
simple, its mathematical study and the proof of Theorem \ref{th:3} is
particularly difficult. 

In the case $\tau=0$, it is possible to get a uniqueness result.
\begin{theo}\label{th:4}{\bf (Uniqueness of the
    solution in the case $\tau=0$; \cite{I})}\\
In the case $\tau=0$, let us assume that the initial data $(\rho^0,\kappa^0)$
is Lipschitz and satisfies for some
$\delta >0$:
$$
\kappa^0_y \ge \sqrt{\delta^2 + (\rho^0_y)^2} \quad \mbox{on}\quad I.   
$$
Then there exists a solution $(\rho,\kappa)$ to
(\ref{eq::12})-(\ref{eq::13})-(\ref{eq::14}) satisfying 
\begin{equation}\label{eq::20}
\kappa_{y}(.,t)\geq \sqrt{\delta^{2}+\rho^{2}_{y}(.,t)} \quad \mbox{for
  all} \quad t>0.
\end{equation}
Moreover this solution is unique among the solutions satisfying
(\ref{eq::20}). 
\end{theo}

\noindent {\bf Main idea for the proof of Theorem \ref{th:4}}\\
The proof of Theorem \ref{th:4} is based on the fact that $\sqrt{\delta^2 +
  \rho_y^2}$ is an entropy subsolution of the conservation law satisfied by
  $v=\kappa_y$, namely
$$v_t=\left(\frac{n(y,t)}{v}\right)_y \quad \mbox{with}\quad
n=\rho_t\rho_y.$$
Then the comparison principle between entropy solutions and entropy subsolutions
shows that $(\rho(\cdot,t),\kappa(\cdot,t))$ satisfies (\ref{eq::20}) for
all time $t>0$.\\

\noindent {\bf Main ideas for the proof of Theorem \ref{th:3}}\\
In particular, a first try shows that
$$M=\kappa_y -|\rho_y|$$
satisfies formally
$$M_t=a_1M_y +a_0M$$
with
$$a_1=\tau \mbox{sgn}(\rho_y) -\frac{\rho_y\rho_{yy}}{(k_y)^2} \quad
\mbox{and}\quad
a_0=\frac{(\rho_{yy})^2}{(\kappa_y)^2}-\frac{\rho_{yyy}\mbox{sgn}(\rho_y)}{\kappa_y}.$$  
By a maximum principle argument, 
we see that in order to guarantee that $M\ge 0$ is true for every time, we
have somehow to control the $L^\infty$-norm of $\rho_{yyy}/\kappa_y$.
This seems hopeless, because we would need to control moreover $\kappa_y>0$
from below. 
The idea is then to replace system  (\ref{eq::12}) by
a suitable regularized system, which is the following  for $\varepsilon>0$:
\begin{equation}\label{eq::17}
\left\{\begin{array}{l}
\displaystyle{\rho_t=(1+\varepsilon)\rho_{yy} -\tau \kappa_y}\\
\displaystyle{\kappa_t=\varepsilon
  \kappa_{yy}+\frac{\rho_{y}\rho_{yy}}{\kappa_{y}}-\tau \rho_{y}}
\end{array}\right.
\end{equation}
and the result is then obtained, passing to the limit $\varepsilon\to 0$.
Even the regularized system (\ref{eq::17}) is still difficult to study,
because of the division by $\kappa_y$ (see \cite{IJM1}). 
But for the regularized system (\ref{eq::17}), it is possible to replace
$M$ by
$$M^\gamma=\kappa_y(\cdot,t) -\sqrt{\gamma^2(t)+(\rho_y(\cdot,t))^2}$$
and to prove that $M^\gamma \ge 0$ while the non-increasing function
$\gamma(t)$ satisfies 
\begin{equation}\label{eq::18}
\displaystyle{\frac{\gamma'}{\gamma}\le -C_\varepsilon\left(1+
    ||\rho_{yyy}(\cdot,t)||_{L^\infty(I)}\right)}   
\end{equation}
where the constant $C_\varepsilon$ blows-up as $\varepsilon\to 0$. The
striking remark is that  to show in the case $\varepsilon>0$ that
$M^\gamma$ is non-negative, 
we only need a control on the $L^\infty$-norm of
$\rho_{yyy}$, while in the case $\varepsilon=0$, in order to show that $M$
is non-negative it was necessary to control  the $L^\infty$-norm of
$\rho_{yyy}/\kappa_y$ which was worse (and not sufficient to conclude).

On the other hand, proving a parabolic version of the well-known
Kozono-Taniuchi inequality (see \cite{KT}) and using heavily the regularity
theory for parabolic equations, we can prove for the regularized system
(\ref{eq::17}) that
\begin{equation}\label{eq::19}
||\rho_{yyy}(\cdot,t)||_{L^\infty(I)} \le Ce^{Ct} (1+|\ln \gamma(t)|)   
\end{equation}
for some constant $C>0$ also depending on $\varepsilon$, among other things.
Putting estimates (\ref{eq::18}) and (\ref{eq::19}) together, we see that
we get the three-exponential estimate
$$\kappa_y(\cdot,t)\ge \gamma(t) \ge e^{-e^{e^{ct}}}$$
for some constant $c>0$ depending  in particular on $\varepsilon$. Then the
global existence of a solution to the regularized system
(\ref{eq::17}) follows. Finally we recover a solution to the original
system (\ref{eq::12}) taking the limit $\varepsilon\to 0$.

\subsection{Numerical simulations}

\begin{figure}[!h]
\begin{center}
\begin{tabular}{lcr}
\epsfig{figure=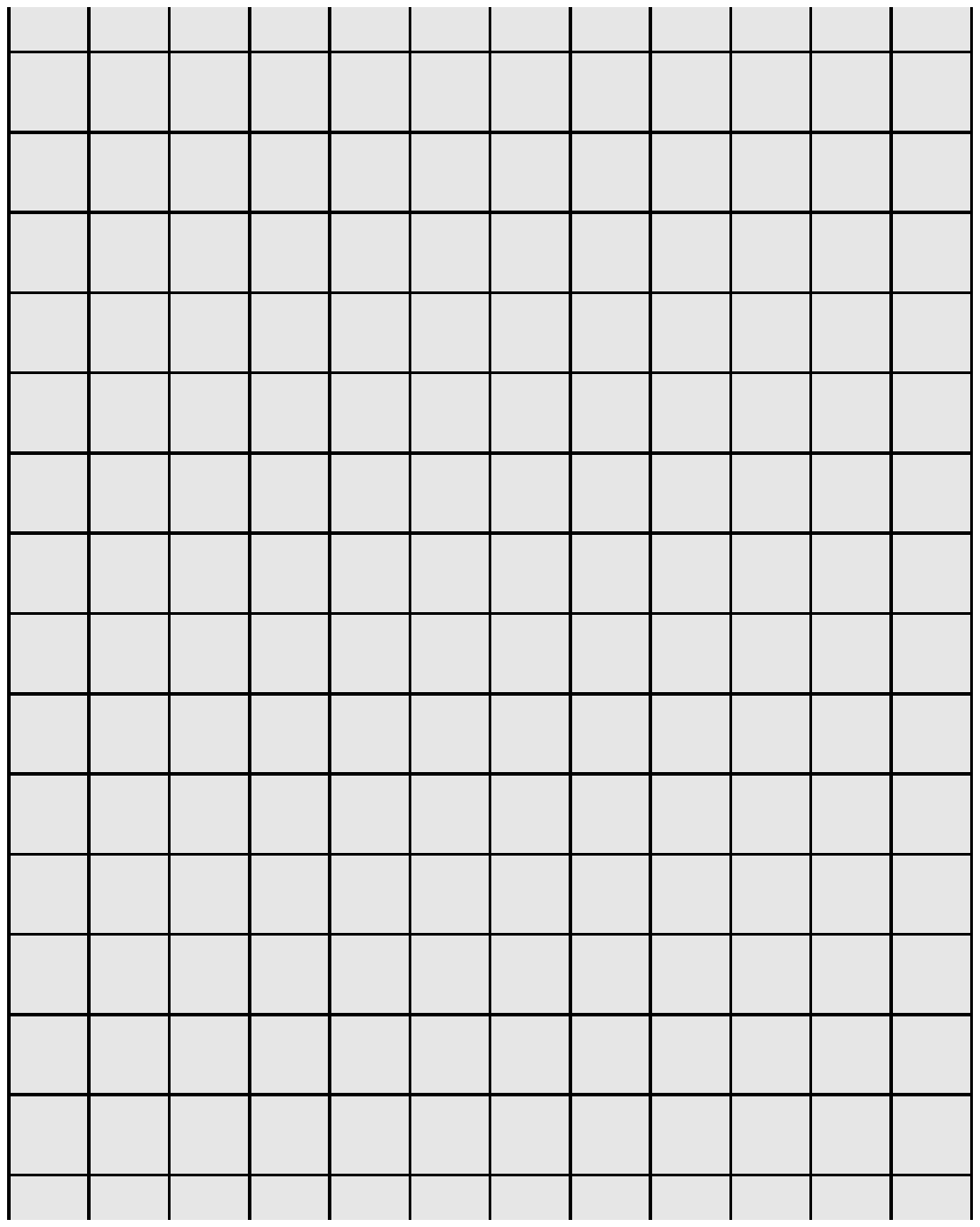,width=25.5mm} 
&$\quad \quad$
\epsfig{figure=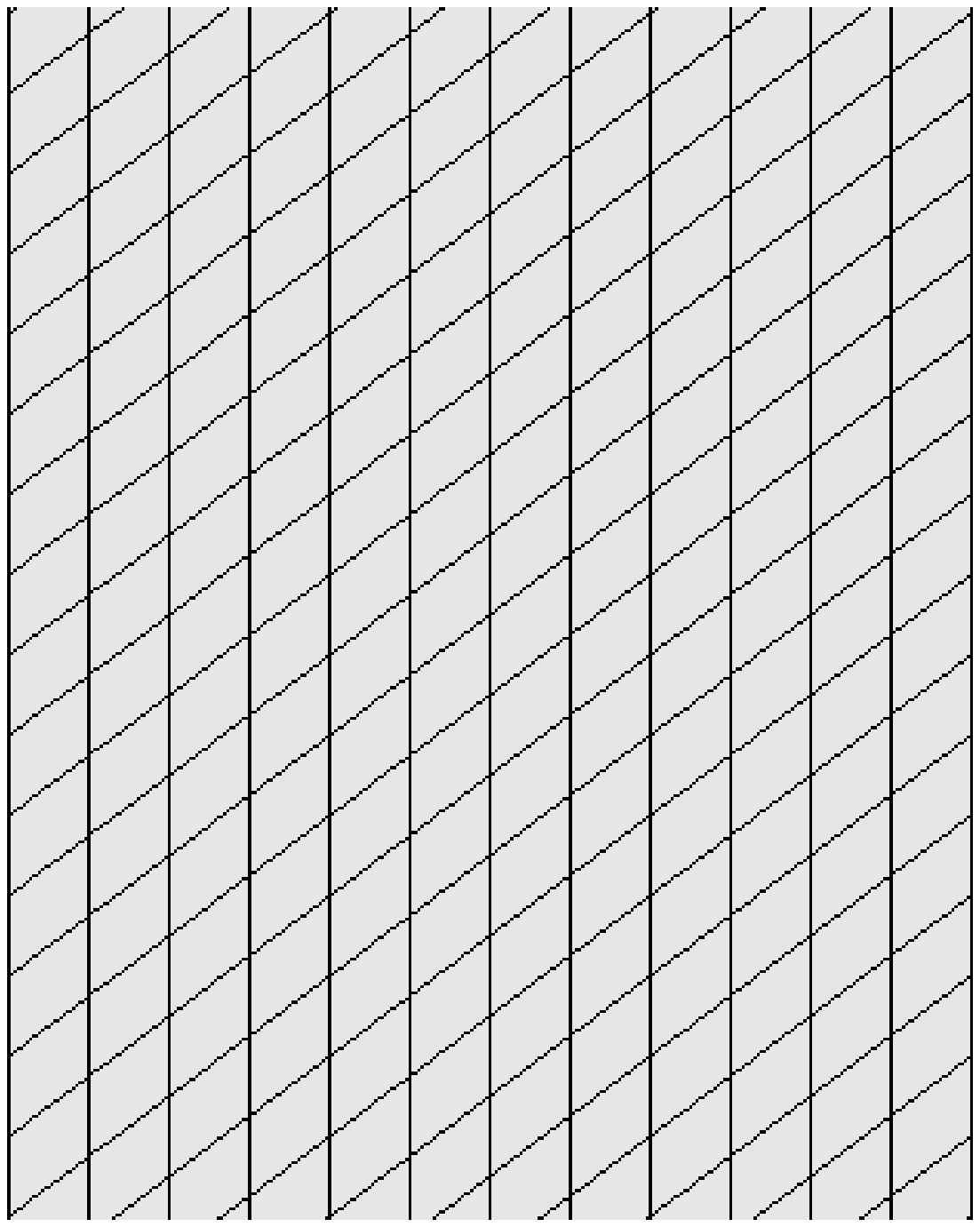,width=25.5mm} 
&$\quad \quad$
\epsfig{figure=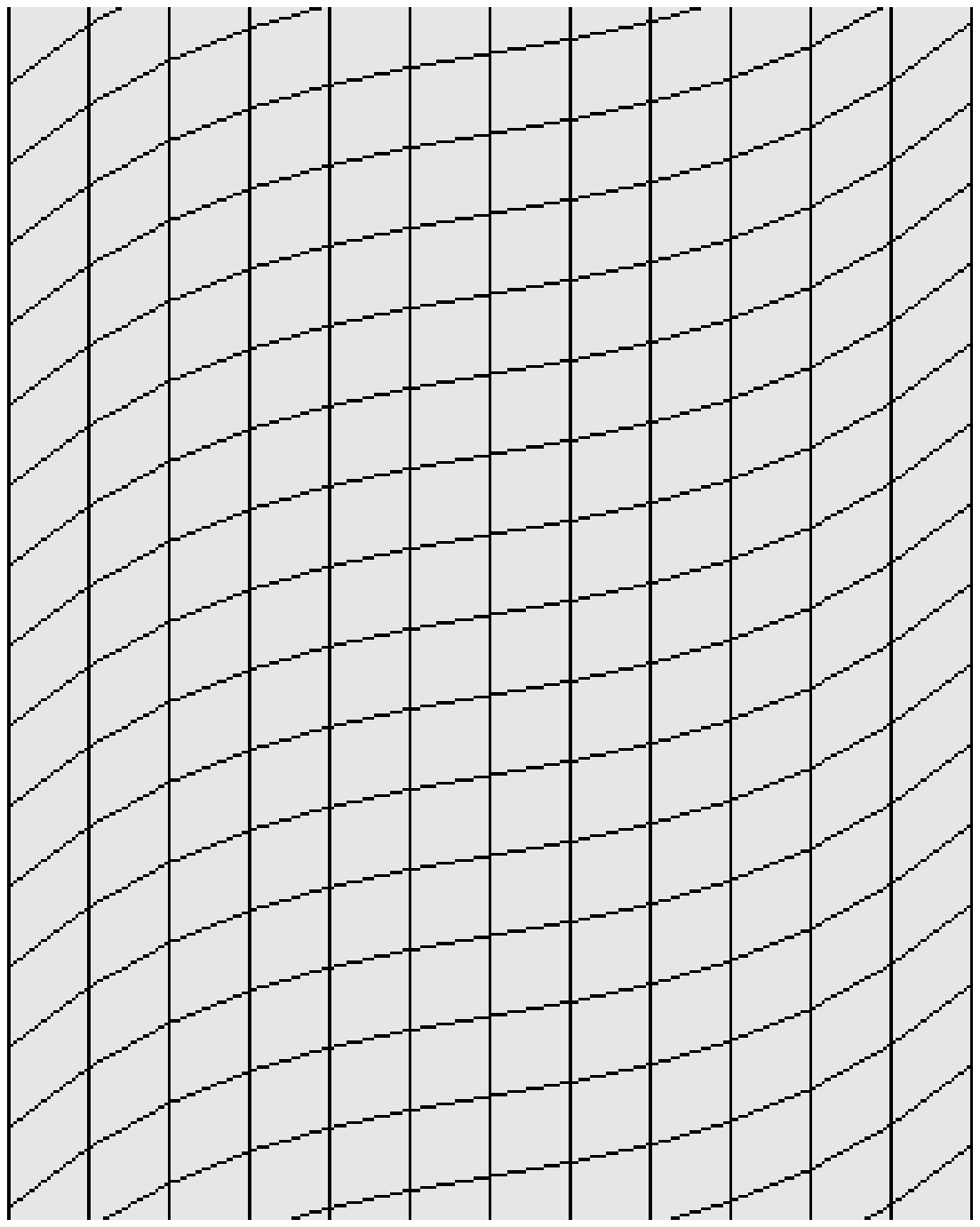,width=25.5mm}\\
$\hspace{0.5cm}$ a) $t=0^{-}$ & $\quad\quad$ b) $t=0^{+}$ & c) $t=+\infty$
$\hspace{0.35cm}$
\end{tabular}
\end{center}
\caption{Deformation of a slab for model (\ref{eq::17})}\label{f5}
\end{figure}

Comming back to the system of elasticity, it is possible to compute the
displacement.
We find $u_1=u_3=0$ and
$$u_2(y,t)=\frac{\tau}{\mu}y  +\int_0^y dz\ \rho(z,t)$$
In Figure \ref{f5}, we show successively the initial state of the crystal
at time $t=0$ without any applied stress, then the instantaneous (elastic) deformation of
the crystal when we apply the shear stress $\tau>0$ at time $t=0^+$. The
deformation of the crystal evolves in time and finally converges
numerically to some particular deformation which is shown on the last
picture after a very long time. This kind of behaviour is called
elasto-visco-plasticity in mechanics.
Moreover, on the last picture, we observe the presence of boundary layer
deformations. This effect is directly related to the introduction of the
back stress $\tau_b$ in the model.
 
\section{General dynamics of curved dislocations}\label{s4}

\subsection{Preliminary on the dynamics of dislocation curves}

\begin{figure}[h]
\centering\epsfig{figure=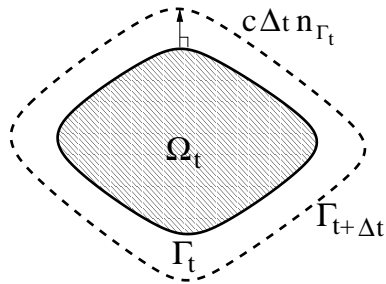,width=50mm}
\caption{Schematic evolution of a dislocation line $\Gamma_t$ by normal
   velocity $c$ between the times $t$ and $t+\Delta t$.}\label{F2}
\end{figure}

At sufficiently low temperature, dislocation curves are contained in the
crystallographic planes of the three-dimensional crystal and can only move in those
planes. Let us consider a closed and smooth curve $\Gamma_t$ moving in the
plane $(y_1,y_2)$. The evolution of this dislocation can be modelled by a
dynamics with normal velocity $c(y,t)$ (see Figure \ref{F2}).

\subsection{Transport formulation of the dynamics}

If we want to describe the dynamics of a high number of dislocations
curves, it is interesting to describe this dynamics with a single quantity
and a single equation. This is the goal of this subsection that we
describe heuristically. 

Let us consider a smooth closed curve $\Gamma_0$ in the plane $(y_1,y_2)$
with basis $(e_1,e_2)$.
The curve $\Gamma_0$ can be parametrized by its curvilinear
abscissa $s$, such that 
$$\Gamma_0= \bigcup_{s} \left\{y(s)\right\}.$$
Let us define for $\theta \in\R/(2\pi \Z)$
$$n(\theta)= (\cos \theta) e_1 + (\sin\theta)e_2 \quad \mbox{and}\quad 
\tau(\theta)=(\sin \theta)e_1-(\cos \theta)e_2.$$
We also define the angle $\theta(s)$ of the tangent to $\Gamma_0$ at the
point $y(s)$ by
$$\frac{dy}{ds}(s)= \tau(\theta(s))$$
and the curvature $K_{\Gamma_0}(y)$ of $\Gamma_0$ at the point $y$ by
$$K_{\Gamma_0}(y(s))=\frac{d}{ds}(\theta(s)).$$
We introduce the lifting $\widehat{\Gamma}_0$ of $\Gamma_0$ by
$$\widehat{\Gamma}_0=\bigcup_{s} \left\{(y(s),\theta(y(s)))\right\}
\quad \quad \subset \quad \quad \R^2\times (\R/(2\pi \Z)).$$
We define the distribution $\delta_{\widehat{\Gamma}_0}(y,\theta)$ on
the test function $\phi(y,\theta)$ by 
$$<\delta_{\widehat{\Gamma}_0},\phi> = \int_{\widehat{\Gamma}_0}ds\ 
\phi(y(s),\theta(y(s))).$$

Let us now consider a smooth evolution of the oriented closed curve $\Gamma_t$ with
normal velocity $c(y,t)$. For any fixed $t$, let us call $s_{\Gamma_t}$ the curvilinear
absissa of the curve $\Gamma_t$ and $s_{\widehat{\Gamma}_t}$ the
curvilinear abscissa  of $\widehat{\Gamma}_t$. Then we define the
distributions
$$g(y,\theta,t)=\left(\frac{ds_{\Gamma_t}}{ds_{\widehat{\Gamma}_t}}\right)\
\delta_{\widehat{\Gamma}_t}(y,\theta)\quad \mbox{and}\quad
\kappa(y,\theta,t)=K_{\Gamma_t}(y) \cdot g(y,\theta,t).$$
We can now state the following result (see Theorem 2.1 and Remark 2.3 in
\cite{M} and the original system for $(g,\kappa)$ in \cite{HZG}):

\begin{theo}\label{th:5}{\bf (Transport formulation of the motion by normal velocity)}\\
Under the previous assumptions, the distributions $g,\kappa$ satisfy the
compatibility equation\\
\begin{equation}\label{eq::21}
\tau\cdot \partial_y  g +\partial_\theta \kappa=0
\end{equation}
and the system
\begin{equation}\label{eq::22}
\left\{\begin{array}{l}
g_t + \partial_y(cng) +\partial_\theta ((\tau \cdot \partial_y c) g)
+c\kappa=0 \\
\kappa_t + \partial_y(cn\kappa) +\partial_\theta ((\tau \cdot \partial_y c)
\kappa) -(n\cdot \partial_y c) \kappa - (\tau\otimes \tau : \partial_{yy} c) g=0.
\end{array}\right.   
\end{equation}   
\end{theo}

Remark that Theorem \ref{th:5} is only known to be true for smooth
evolutions of closed curves. When singularities appear (in general in
finite time), we do not know if system (\ref{eq::22}) is still true.

\subsection{Dynamics of densities of dislocation curves}

We now consider the three dimensional case with the basis $(e_1,e_2,e_3)$.
Let us now not only consider dislocations in the particular plane $(y_1,y_2)$ for
$y_3=0$, but also in parallel planes for any $y_3=\mbox{constant}$. To this
end, let us consider distributions
$g(y,\theta,t), \kappa(y,\theta,t)$ with  the new notation
$y=(y_1,y_2,y_3)$, which are again assumed to satisfy (\ref{eq::21}) and
(\ref{eq::22}).
To close the system we have to explain how is defined the normal velocity
$c(y,t)$ in the case of dislocation dynamics. To this end, we have first to
compute the strain $e(y,t)\in \R^{3\times 3}_{sym}$ which is solution of
the system
\begin{equation}\label{eq::23}
\left\{\begin{array}{l}
\mbox{div}\ \sigma=0 \quad \mbox{with}\quad \sigma= 2\mu e + \lambda
(\mbox{trace}(e)) Id\\
\mbox{inc}\ e=(\mbox{curl}_{row}\ (b\times \beta))_{sym}\quad
\mbox{with}\quad \beta(y,t)=\int_{\R\backslash (2\pi \Z)} d\theta \ \tau(\theta) g(y,\theta,t)
\end{array}\right.   
\end{equation}
where the operator $\mbox{inc}\ e$ is obtained, taking first the curl of
the column vectors of the matrix $e$, and then the curl of the row vectors
of the new matrix. The $\mbox{curl}_{row}$ is the curl of the row vectors
of the matrix, and the index $(\ )_{sym}$ means that we consider the
symmetric part of the matrix. The quantity $b\otimes \beta$ is called the
Nye tensor of dislocation densities, where $b$ is the Burgers vector of the
dislocations under consideration. Then, to close the model, we define the
normal velocity field by 
\begin{equation}\label{eq::24}
c(y,t)=(b\otimes e_3):\sigma(y,t).   
\end{equation}
Relation (\ref{eq::24}) between the normal velocity and the right hand side
(called the resolved Peach-Koehler force) is the simplest expression when
the drag coefficient is isotropic and taken equal to $1$.
Again, if we neglect the short range dynamics, the complete system is
(\ref{eq::21})-(\ref{eq::22})-(\ref{eq::23})-(\ref{eq::24}), and can be
interpreted as a mean field model, describing the densities of dislocation
curves moving in interaction.\\

Finally, remark that the Groma, Balogh system (\ref{eq::1}) is a particular
case of this complete model, when every quantity is invariant in $y_2$ and
$x_1=y_1$, $x_2=y_3$, with the notation
$\theta^\pm(x_1,x_2,t)=g(y_1,0,y_3,\pm \pi,t)$.

\section{Concluding remarks}\label{s5}

The models presented in this paper deal with the dynamics of dislocation densities, and
can be seen as a very first (and naive) attempt to derive
elasto-visco-plasticity behaviour of crystals. All these models are
essentially mean field models, and for instance are not able to explain
relations between the stress and the plastic deformation velocity like for
instance the typical power-law behaviour
$$\dot{e}^p \simeq \pm |\sigma|^m$$
with $m$ large.
More realistic models should also take into
account the complicated short range dynamics, with the possible pinning of
dislocations, the creation and destruction of dipoles of dislocations with
opposite Burgers vector, and even the possibilities of junctions between
dislocations, or the Frank-Read sources and the cross-slip phenomenon.
At least, it seems that the framework of kinetic equations for the creation and
annihilation of such dipoles should be a promising tool for the modelling
of the short range dynamics of dislocations. We hope to investigate this
modelling in a future work.

\section*{Acknowledgement}
This work was supported by the contract ANR MICA (2006-2009). The
authors thank the organizers of the meeting CANUM 2008, for the
apportunity to present recent results on dislocation dynamics.

\end{document}